\documentclass[cup7b]{cupbook}

\usepackage{graphicx}
\usepackage{amsmath,amssymb}

\newcommand{\C}{\mathbb{C}}
\newcommand{\bp}{\mbox{\boldmath $p$}}
\newcommand{\invg}{\mathcal{G}}
\newcommand{\hil}{\mathcal{H}}
\newcommand{\mani}{\mathcal{M}}

\begin{document}
\author[O.~Dreyer]{Olaf Dreyer\\
Theoretical Physics, Blackett Laboratory,\\
Imperial College London, London, SW7 2AZ, U.K.}
\chapter{Emergent General Relativity}
\section{Introduction}
\begin{figure}
\begin{center}
\includegraphics[height=3.5cm]{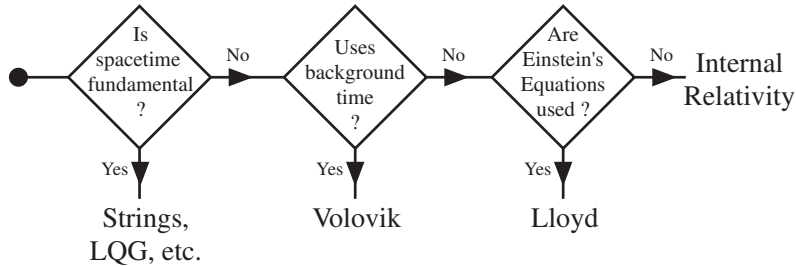}
\end{center}
\caption{Choices on the road to quantum gravity.}
\label{fig:order}
\end{figure}

This article wants to be two things. On the one hand it wants to review a number of approaches to the problem of quantum gravity that are new and have not been widely discussed so far. On the other hand it wants to offer a new look at the problem of quantum gravity. The different approaches can be organized according to how they answer the following questions: Is the concept of a spacetime fundamental? Is a background time used? Are Einstein's equations assumed or derived? (See figure \ref{fig:order}). 

In string theory, loop quantum gravity, and most other approaches reviewed in this book spacetime plays a fundamental role. In string theory a given spacetime is used to formulate the theory, in loop quantum gravity one tries to make sense of quantum superpositions of spacetimes. It is these spacetimes in the fundamental formulation of the theory that are directly related to the spacetime we see around us. In this broad sense these approaches treat spacetime as something fundamental. Here we want to focus our attention on approaches that take a different view. In these approaches spacetime emerges from a more fundamental theory.

The next questions concern the role of time. The models that we will be looking at will all have some sort of given time variable. They differ though in the way they treat this time variable. One attitude is to use this time variable in the emergent theory. The goal of quantum gravity in this context could then be to find a massless spin two particle in the excitation spectrum of the Hamiltonian corresponding to the given time. We will see in section \ref{ssec:volovik} a solid state physics inspired approach due to G.~Volovik that takes this point of view.

The other possible attitude towards the background time is that it is just a fiducial parameter that is not important for the emergent physics. If one takes this view then there is one more question: What is the role of the Einstein equations? In section \ref{ssec:lloyd} we will see a quantum information theory inspired model by S.~Lloyd that \emph{uses} the Einstein equations to formulate the theory. The other possibility is to argue for why the Einstein equations hold true. In section \ref{sec:intrel} we will show how such an argument can be made. We call this approach Internal Relativity. 

\section{Two views of time}
In this section we review to approaches to quantum gravity that differ in the way they view time. The first approach comes from solid state physics; the second comes from quantum information theory. 

\subsection{Fermi points}\label{ssec:volovik}
\begin{figure}
\begin{center}
\includegraphics[height=8cm]{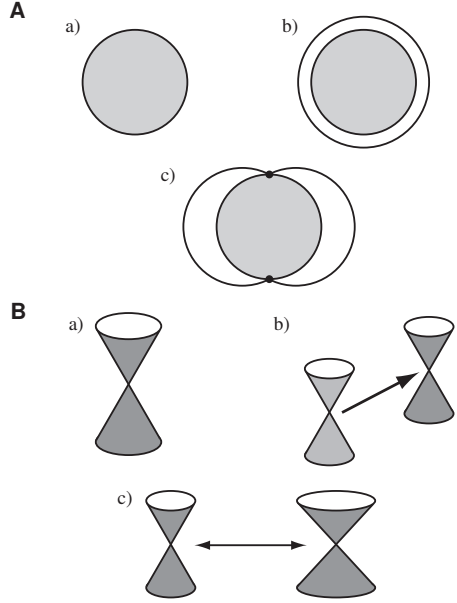}
\end{center}
\caption{\textbf{\textsf{A}} Possible momentum space topologies for a Fermi liquid. a) a Fermi surface, b) a fully gapped system, and c) a system with Fermi points. \textbf{\textsf{B}} The possible excitations of a system with Fermi points. a) The light cone of the emergent fermions, b) moving the Fermi points corresponds to gauge degrees of freedom, c) shape changes of the light cone give a kind of emergent gravity.}
\label{fig:volovik}
\end{figure}

In this section we are interested in the low energy behavior of quantum mechanical Fermi liquids. It turns out that this behavior does not depend on the details of the model but is rather described by a small number of universality classes. Which universality class a given model falls into is determined by the topology of the energy spectrum in momentum space. The best known class is that of a simple Fermi surface (see figure \ref{fig:volovik}\textbf{\textsf{A}}a). In an ideal free Fermi gas the Fermi surface is the boundary in momentum space between the occupied and unoccupied states. If $p_F$ is the corresponding momentum then the energy spectrum is given by
\begin{equation}
E( \bp ) = v_F (\vert\bp\vert - p_F).
\end{equation}
In addition to these fermionic degrees of freedom there are also bosonic excitations given by oscillations of the Fermi surface itself. The dynamics of the fermionic and bosonic degrees of freedom is described by the Landau theory of Fermi liquids. 

The other well known situation is that of a fully gapped system (see figure \ref{fig:volovik}\textbf{\textsf{A}}b). In this case the next available energy level above the Fermi surface is everywhere separated from it by a non-zero amount $\Delta$. This situation is encountered in superfluids and superconductors. 

Most interesting for us is the situation when the gap $\Delta$ is not uniform but vanishes at certain points (see figure \ref{fig:volovik}\textbf{\textsf{A}}c). These points are called \emph{Fermi points}.    It is the low energy behavior of this universality class that shows the kind of excitations we see around us: Fermions, gauge fields, and even gravity. This happens because a Fermi point is a stable feature that is insensitive to small perturbations (see \cite{volovik, horava} for more details). Its presence is protected by topology. The Fermi point itself represents a singularity in the Fermi propagator $G$. Its inverse $\invg$ has a zero at the Fermi point. If we think of a small sphere $S^3$ centered at the Fermi point then $\invg$ defines a map  
\begin{equation}
\invg : S^3 \longrightarrow \text{GL}(N, \C),
\end{equation}
where $N$ is the number of components of the fundamental fermions including internal indices. Thus $\invg$ defines an element in $\pi_3(\text{GL}(N, \C))$, the third homotopy group of $\text{GL}(N, \C)$. If this homotopy class is non-trivial the Fermi point can not be removed by a small perturbation. 

Since the inverse propagator $\invg$ vanishes at the Fermi point it has the following expansion near the Fermi point:
\begin{equation}
\invg(p) = \sigma^a e_a^\mu (p_\mu - p^{0}_\mu),
\end{equation}
where we have for concreteness assumed that we have two spin components so that the Pauli matrices $\sigma^a$, $a=0,\ldots, 3$, can be used as a basis. Given that the Fermi points can not disappear the effect small perturbations can have is rather restricted. It can move the position of the Fermi point (see figure \ref{fig:volovik}\textbf{\textsf{B}}b) or it can change the shape of the light cone (see figure \ref{fig:volovik}\textbf{\textsf{B}}c). The parameters $e_a^\mu$ and $p^{0}_\mu$ appearing in the above expansion thus become dynamic. We can infer the physical meaning of these new dynamical degrees of freedom by looking at the energy spectrum. The spectrum is determined by the zero of $\invg$. Here we obtain 
\begin{equation}
g^{\mu\nu} (p_\mu - p^{0}_\mu)(p_\nu - p^{0}_\nu) = 0,
\end{equation}
where
\begin{equation}
g^{\mu\nu} = \eta^{a b} e_a^\mu e_b^\nu.
\end{equation}
The change in the shape of the light cone can thus be identified with a changing metric $g^{\mu\nu}$ and the change of the position of the Fermi point gives rise to an electromagnetic field $A_\mu$:
\begin{equation}
A_\mu = p^0_\mu.
\end{equation}
We thus see that the low energy physics of a Fermi liquid with a Fermi point possesses all the kind of excitations that we see around us, i.e. fermions, gauge fields, and dynamics. Unfortunately the mass of the graviton is not generically zero. Instead the parameters of the model have to be chosen in a special way to make the mass vanishingly small. 

\subsection{Quantum computation}\label{ssec:lloyd}
A completely different approach is the one proposed by S.~Lloyd \cite{lloyd}. For him the universe is one giant quantum computation. The problem of quantum gravity is then to show how a quantum computation gives rise to a spacetime. 

\begin{figure}
\begin{center}
\includegraphics[height=8cm]{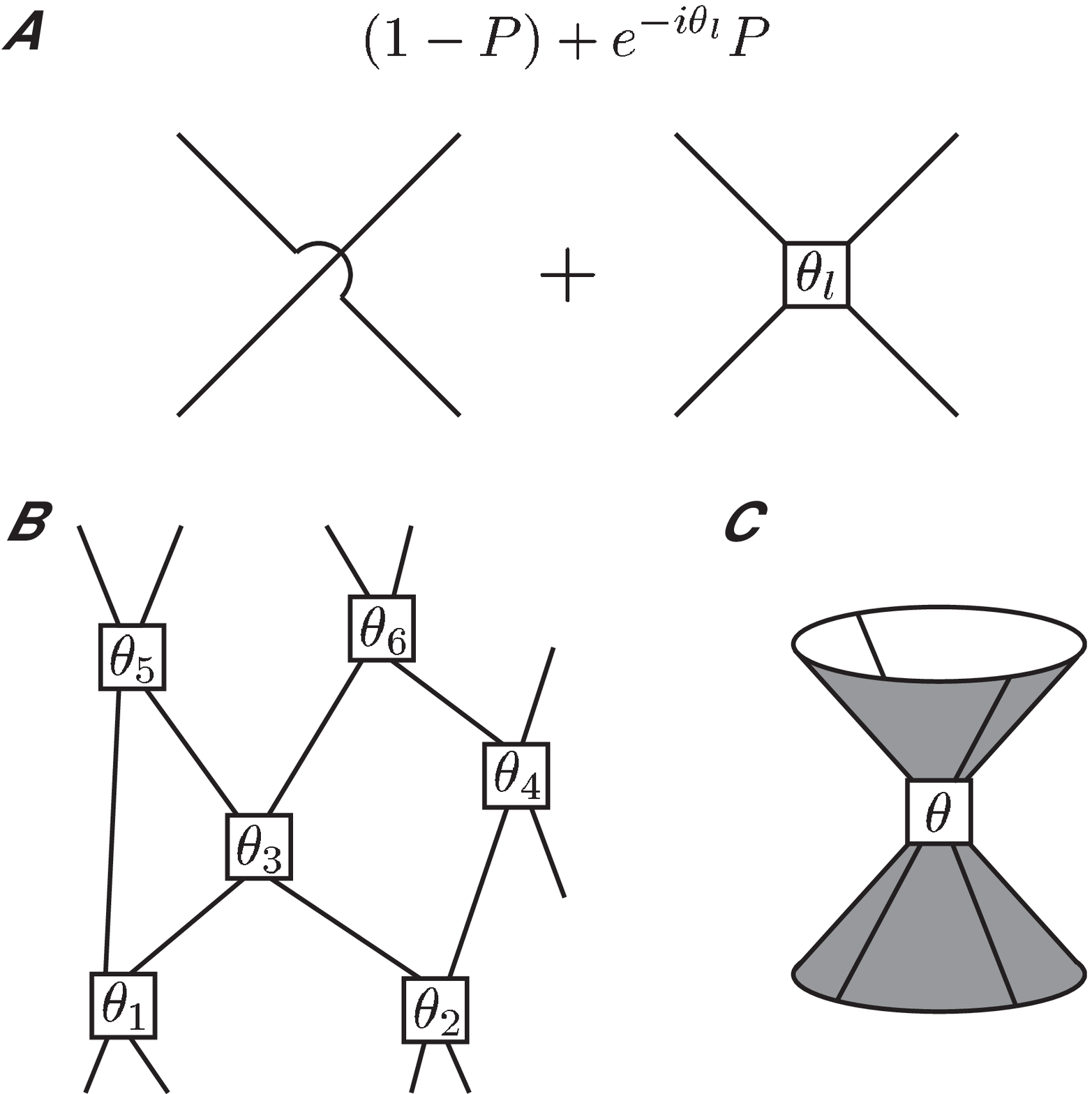}
\end{center}
\caption{\textbf{\textsf{A}}  The unitary $U_l$ consists of two parts. On the right the two qubits scatter off each other, giving rise to a phase $\theta_l$. On the right the qubits miss each other. \textbf{\textsf{B}} The different $U_l$, $l=1,\ldots,n$, give rise to $2^n$ different possible computational histories. Each history consists of a causal set and a set of phases $\theta_l$. \textbf{\textsf{C}} The two incoming and outgoing qubits give four directions on the light cone at a node of the causal set. These four null directions determine four of the ten components of the metric.}
\label{fig:lloyd}
\end{figure}

A quantum computation is given by a unitary operator $U$ acting on the Hilbert space of our system. Here we take this system to be $N$ qubits. The Hilbert space is thus 
\begin{equation}
\hil = (\C^2)^{\otimes N}.
\end{equation}
We can decompose $U$ into quantum gates $U_l$, $l=1, \ldots, n$, that are acting on two qubits at a time:
\begin{equation}
U = U_n \cdots U_1
\end{equation}
It is here that a discrete form of background time makes its appearance. The individual $U_l$'s appear in a definite order given by the parameter $l$. We will see though that this time is not related to the time as perceived by an observer in the model. Without restriction we can assume that the $U_l$'s have the form
\begin{eqnarray}
U_l & = & e^{-i\theta_l P}\\
 & = & (1 - P) + e^{-i\theta_l} P,\label{eqn:scatter}
\end{eqnarray}
for a projection operator $P = P^2$. We can represent such $U_l$'s as in figure \ref{fig:lloyd}\textbf{\textsf{A}}. The two parts of equation \ref{eqn:scatter} can be given a physical interpretation. In the subsystem that $P$ projects onto the two qubits that $U$ is acting on scatter. This results in a phase shift of $\theta_l$. In the orthogonal subspace the two qubits do not scatter. Here there is no phase shift. The whole unitary $U$ can now be written as follows
\begin{eqnarray}
U & = & \prod_{l=1}^n ((1-P_l) + e^{-i\theta_l}P_l)\\
  & = & \sum_{b_1, \ldots, b_n \in \{0,1\}}e^{-i\sum_{l=1}^n b_l\theta_l}P_n(b_n)\cdots%
  P_1(b_1), \label{eqn:causalsum}
\end{eqnarray}
where we have denoted $(1-P_l)$ by $P_l(0)$ and $P_l$ by $P_l(1)$. Given our interpretation of the individual $U_l$'s we can represent the terms in the sum (\ref{eqn:causalsum}) as causal sets as in figure \ref{fig:lloyd}\textbf{\textsf{B}}. We will call such a causal set together with the angles $\theta_l$ a computational history. Any quantum computation can thus be interpreted as a superposition of computational histories. 

The next step is to interpret each computational history as a discrete spacetime. To see this we embed all the histories into one manifold $\mani$. The lines of the calculation that  run between the scattering events are identified with null geodesics of the metric. At each node of the causal set we have four vectors that lie on the light cone at that point (see figure \ref{fig:lloyd}\textbf{\textsf{C}}). It follows that at this point four of the ten components of the metric are given. To find the remaining six we use the Einstein equations in their Regge form. Before we can use Regge calculus we have to turn our causal sets into simplicial lattices. The added lines will in general no longer be null. The metric will be fully determined once the lengths of all these additional lines are specified. 

We will choose the lengths of the new lines in such a way that the Einstein equations
\begin{equation}
\frac{\delta I_G}{\delta g} + \frac{\delta I_M}{\delta g} = 0,
\end{equation}
are satisfied. Here $I_G$ is the gravitational action in its Regge calculus form and $I_M$ is the matter action which is a function of the $\theta_l$'s and the metric (i.e. the length of the lines). 
	
Given a quantum computation we arrive at a superposition of discrete spacetimes. Since this is a quantum superposition one still has to argue how the classical limit is achieved. Note though that the task is easier in this setup since all the computational histories are embedded into one manifold $\mani$. There is no problem in identifying points in the different histories as there is in other approaches to quantum gravity.

One problem that remains is the universality of the above construction. We have assigned spacetimes to all quantum computations. It is not clear what the meaning of this spacetime picture is for a generic quantum computation. The question arises of what the right calculation is. 

\section{Internal Relativity}\label{sec:intrel}
In the last section we have encountered two approaches to quantum gravity in which the metric emerges. In the view proposed by Volovik gravity emerges as a massless spin two excitation of a Fermi system with a Fermi point. We are able to find gravity using the background time the theory is formulated in. In the computational universe there is also a background time but it plays no role in the spacetime constructed from the computational histories. In this construction the Einstein equations are used. The approach we want to  propose here is similar in that only internally available information is used to reconstruct a spacetime. It differs in that the Einstein equations are not used but are to be derived.

\subsection{Manifold matter}
The most important ingredient in our construction are coherent degrees of freedom. It is these degrees of freedom that provide the glue that makes the manifold. Without them there is no notion of causality. Given two such coherent degrees of freedom we can identify a point by the intersection of the two. Our manifold will consist of points of this kind. 

An example of coherent degrees of freedom is provided by a simple spin model from solid state physics. The XY-model is given by the Hamiltonian 
\begin{equation}
H = \sum_{i=1}^N (\sigma^+_i\sigma^-_{i+1} + \sigma^-_i\sigma^+_{i+1}),
\end{equation}
where $\sigma^\pm = \sigma^x \pm i \sigma^y$, and the $\sigma$'s are the Pauli matrices. This model can be solved exactly using the Jordan-Wigner transformation \cite{jordanwigner}. One obtains a free fermionic model described by the Hamiltonian
\begin{equation}
H = \sum_{k=1}^N\varepsilon(k) f_k^\dagger f_k,
\end{equation}
where $f$, and $f^\dagger$ are the annihilation and creation operators for the fermions and $\varepsilon(k)$ is the energy
\begin{equation}
\varepsilon(k) =  16\pi \cos \frac{2\pi}{N} k. 
\end{equation}
One ground state can be obtained by half filling the Fermi sea. The excitations then have a linear dispersion relation given by
\begin{equation}
\Delta\varepsilon = 16\pi J_\perp \frac{2\pi}{N}\Delta k\equiv v_\text{F} \Delta k. 
\end{equation}
It is excitations like these that play the role of our coherent degrees of freedom. The above example is too simple to stand in as a model for our world. A far more interesting example has recently been proposed by X.-G.~Wen \cite{wen}. Although it is also built with simple spins it has both fermions and gauge interactions in its low energy limit. The particles of this model make for far more interesting coherent degrees of freedom that we can use in our construction.

Compare this notion with what we have seen in the computational universe. The coherent degrees of freedom are the lines in the computational graph, i.e. the qubits, and the points are the places they interact, i.e. the quantum gates. Compare also the article by F. Markopoulou in this volume in which coherent degrees of freedom are described by noiseless subsystems, a quantum information theoretic notion. Thus we have the following correspondences:
\begin{center}
coherent degree of freedom\\
$\equiv$\\
noiseless subsystem\\
$\equiv$\\
qubits in computational history
\end{center}
The correspondences for the points of the manifold are thus:
\begin{center}
points of manifold\\
$\equiv$\\
intersections of coherent degrees of \\
freedom/noiseless subsystems\\
$\equiv$\\
quantum gates
\end{center}
We want to stress one point here that all the proposals so far have in common. For all of them spacetime and matter arise together. They can not be separated. We will see in section \ref{sec:cons} that it is here that the cosmological constant problem solved. 

\subsection{Metric from dynamics}
Having introduced our manifold as the set of coincidences of coherent degrees of freedom we now want to endow this set with a metric structure. How are we going to go about this? It is clear that there is one thing that we can not do. We can not use the background structure to introduce notions of distance or time. That means that the lattice our theory is defined on and the background time can not be used for this purpose. Instead what we will have to do is to use only notions that are internally available in the system. These are again our coherent degrees of freedom.

In our current system of units we are using light and cesium atoms to define what we mean by space and time. In the language used in this article we would say that we are using coherent degrees of freedom to arrive at metric notions. In our above example a spin wave could play the role that is played by light for us. Since we only allow access to such internal information it is not hard to see that the world will look relativistic to observers in the system.

\begin{figure}
\begin{center}
\includegraphics[height=4cm]{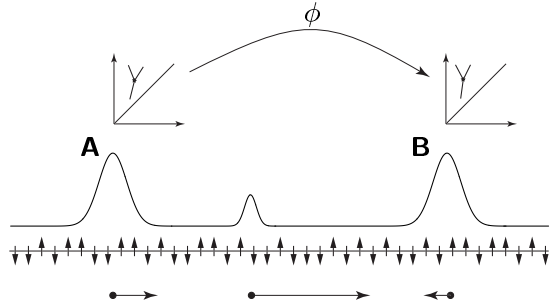}
\end{center}
\caption{ A view of the system that is not available to
  observers confined inside the system. The observers
  \textbf{\textsf{A}} and \textbf{\textsf{B}} have no way of telling what
  their motion is with respect to the lattice. This is why 
  both observers assign the same speed to
  the excitation. There exists a map $\phi$ between the two
  coordinate systems given by the mapping of physical events onto each other.
  This map $\phi$ will have the property that it maps light onto light.
  We find then that this map $\phi$ must be a Poincar\'e transformation.}
\label{fig:poincare}
\end{figure}

Since the observers have no access to the underlying model they can not tell whether they are moving with respect to it. They will thus all assign the same speed to the coherent degree of freedom. The only transformation between their respective coordinate representations is then a Poincar\'e transformation since this is the only transformation that leaves the speed of the excitations unchanged (see figure \ref{fig:poincare}). 

It is in this sense that our approach is background independent. It is only through the dynamics of the system and the emergent coherent degrees of freedom that we arrive at metric notions. 

This is again to be compared with the computational universe. The qubits are defined to be null just as the coherent degrees of freedom in our approach are null. The same is true for the noiseless subsystems of F. Markopoulou. 

\subsection{The equivalence principle and the Einstein equations}
We now want to look at why our approach includes more then just flat Minkowski space. Having introduced metric notions we now want to proceed to define notions like mass and energy. It is here that we will see that the presence of a massive body will have an influence on the spacetime surrounding it. 

When defining the mass of a body we have to do it in an internal or background independent way. One such way was described by E.~Mach \cite{mach}. If one takes two masses $m_1$ and $m_2$ and makes them collide the changes in velocity $\Delta v_1$ and $\Delta v_2$ will be related by
\begin{equation}
\frac{m_1}{m_2} = - \frac{\Delta v_2}{\Delta v_1}.
\end{equation}
Given one standard mass this relation can be used to define all other masses. Note that this definition of inertial mass is completely relational. Note also that for this definition to work the theory can not be free. We need interactions for the two masses to bounce off of each other. It is here that things become interesting. To have a notion of mass for our coherent degrees of freedom they have to interact. But it is these same degrees of freedom that we have used to define our notions of spacetime. That means that the metric changes because of the presence of a massive object. 

This connection between inertial and gravitational mass is well known and goes under the name of equivalence principle. We have argued that the equivalence principle follows from background independence. We want to go one step further and make the following conjecture:

\begin{description}
\item[Conjecture] When notions of distance, time, mass, energy, and momentum are defined in a completely internal way the Einstein equations hold.
\end{description}

Let us call this approach to the problem of quantum gravity \emph{internal relativity} to stress the internal background independent point of view. 

\subsection{Consequences}\label{sec:cons}
Our point of view sheds light on two long standing puzzles: the cosmological constant problem and the problem of time. Here we want to describe shortly how these problems dissolve when spacetime and matter are not treated separately.

This cosmological problem arises when one views quantum field theory as a theory describing fields living on a curved spacetime. This view runs into a serious problem when one considers the effect the quantum fields should have on spacetime. Since all the modes of the quantum field have a zero energy of $\pm 1/2\hbar\omega$, one expects a contribution to the vacuum energy on the order of 
\begin{equation}
\int^\Xi d\omega\;\hbar\omega^3 \sim \hbar\,\Xi^4,
\end{equation}
where $\Xi$ is some high energy cut-off. If one takes this cut-off to be the Planck energy the vacuum energy is some 123 orders of magnitude away from the observed value of the cosmological constant, making this the worst prediction in theoretical physics. 

We see that the root of the cosmological constant problem lies in the fact that we have treated spacetime and matter as separate objects. If we treat quantum fields as living on a spacetime, then we will encounter the cosmological constant problem. If, on the other, hand we realize that it is only through the excitations described by the quantum fields that a spacetime appears in the first place, the above argument can not be given and the cosmological constant problem disappears. 

The problem of time appears when one tries to quantize the gravitational field on its own. Because the Hamiltonian vanishes there is no notion of time evolution left. In our approach it does not make sense to treat the gravitational field without matter. To do so means stepping into the ``problem of time" trap.

\section{Conclusion}\label{sec:concl}
In this paper we have tried to review a number of approaches to the problem of quantum gravity in which spacetime is emergent. We have seen that even when spacetime is not fundamental there are still a number of choices to be made. The first choice to be made concerns the role of time. Is the background time to be used or is it more like a fiducial parameter? 

An example were the background time is used is Volovik's theory of Fermi Liquids with a Fermi point. The quest here is for a theory that has a massless spin two particle in its spectrum. We have seen that Volovik comes close. It is the mass of the graviton that is the problem. Generically it will not vanish. It is interesting though that this model reproduces a lot of the physics we see around us, including fermions and gauge excitations. 

As an example where the role of time is different we have seen Lloyd's computational universe. The discrete time labeling the individual quantum gates $U_l$, $l=1, \ldots,n$, is not used in the construction of the spacetime metrics of the computational histories. Note how the questions changes here. One is no longer looking for a massless spin two excitation. In the context where the whole spacetime metric is to be defined it would not even be clear what a massless spin two excitation would mean. Instead one looks for the whole metric using the Einstein equations. 

This attempt is also not without problems. Given \emph{any} quantum computation one can construct computational histories with a corresponding spacetime interpretation. The question of the meaning of these metrics then arises. Why is there a spacetime interpretation to a calculation that factorizes large integers?

The proposals reviewed here were all presented at a workshop at the Perimeter Institute in Canada\footnote{Recordings of the talks can be found on the website of the Perimeter Institute at www.perimeterinstitute.ca.} We have not discussed approaches that are included in this volume through the contributions of participants to the workshop. See R.~Loll, F.~Markopoulou, and the string theorists that have also ventured into the realm of emergent spacetime. 

In addition to the proposals presented at the workshop we have also discussed a novel approach which differs from the computational universe mainly in that it does not use the Einstein equations. We instead argued that they are a result of the internal and background independent approach.

The main ingredient are coherent degrees of freedom. These play the role of matter but they are also used to define notions of space and time. It is because they play this dual role that the equivalence principle and also the Einstein equations are true. 

In this approach there is no notion of spacetime without matter. Tearing apart spacetime and matter by viewing the latter as living on the former creates deep problems like the cosmological constant problem and the problem of time. Here we avoid these problems.

This view also goes well with a new view of quantum mechanics \cite{dreyerqm}. In this view of quantum mechanics a notion like position is only applicable to large quantum systems and is not fundamental. Given such a view, it is only natural that a spacetime emerges and is not included as a basic building block. 

In recent years we have see a number of new approaches to the problem of quantum gravity come very close to the stated goal. Using methods and ideas foreign to the more traditional approaches they were able to make progress were others got stuck. Maybe we will soon have not just one quantum theory of gravity but several ones to choose from. To decide which one is the right one will then require recourse to experiment. What an exciting possibility.

\end{document}